# The Origin of Ferroelectricity in $Hf_xZr_{1-x}O_2$: A Computational Investigation and a Surface Energy Model

R. Materlik[1)], C. Künneth[1)] and A. Kersch [1,a)]

[1] *Munich University of Applied Sciences, Department of Applied Sciences and Mechatronics, Lothstr. 34, 80335 Munich, Germany*

---

[a)] *Author to whom correspondence should be addressed. Electronic mail: akersch@hm.edu.*

The structural, thermal, and dielectric properties of the ferroelectric phase of $HfO_2$, $ZrO_2$ and $Hf_{0.5}Zr_{0.5}O_2$ (HZO) are investigated with carefully validated density functional computations. We find, that the free bulk energy of the ferroelectric orthorhombic $Pca2_1$ phase is unfavorable compared to the monoclinic $P2_1/c$ and the orthorhombic Pbca phase for all investigated stoichiometries in the $Hf_\chi Zr_{1-\chi}O_2$ system. To explain the existence of the ferroelectric phase in nanoscale thin films we explore the Gibbs / Helmholtz free energies as a function of stress and film strain and find them unlikely to become minimal in HZO films for technological relevant conditions. To assess the contribution of surface energy to the phase stability we parameterize a model, interpolating between existing data, and find the Helmholtz free energy of ferroelectric grains minimal for a range of size and stoichiometry. From the model we predict undoped $HfO_2$ to be ferroelectric for a grain size of about 4 nm and epitaxial HZO below 5 nm. Furthermore we calculate the strength of an applied electric field necessary to cause the antiferroelectric phase transformation in $ZrO_2$ from the $P4_2/nmc$ phase as 1 MV/cm in agreement with experimental data, explaining the mechanism of field induced phase transformation.

## I. INTRODUCTION

The recent discovery of ferroelectricity in $HfO_2$ and $ZrO_2$ based high-k materials[1,2] surprised since this material class was extensively studied for decades[3,4]. Similar to the classical perovskites, the origin of the ferroelectricity is most likely a non-centrosymmetric polar crystal phase which is stable under certain conditions. Kisi[5] found and structurally analyzed the polar orthorhombic $Pca2_1$ phase (f-phase) more than 2 decades ago in $ZrO_2$ under asymmetric stress conditions. In this time theoretical calculations were performed[6] showing such a phase to be stable. Neither the dielectric properties were studied experimentally nor examined by calculations. The discovery of the ferroelectric properties arised from the development of nanoscale thin films[7] for use in memory technology, where crystalline doped films of high quality were developed. After the observation of a remanent polarization and hysteresis in 10 nm thick films of 3 % Si doped $HfO_2$ films[1] in a TiN-$HfO_2$-TiN stack, the effect was additionaly found in Al- and Y-doped films[8,9] grown on TiN as well. More suitable dopants have been discovered since[10]. Furthermore, ferroelectricity was found in the 9 nm mixed oxide thin film $Hf_{0.5}Zr_{0.5}O_2$ (HZO)[2] deposited on TiN electrodes, although neither the pure $HfO_2$ nor the pure $ZrO_2$ film exhibit ferroelectricity. The existence of a true ferroelectric effect has been confirmed by various experiments such as P-V and C-V measurements [8] as well as retention extrapolated up to 10 years [12] and the occurrence of orthorhombic peaks in XRD measurements [13]. $ZrO_2$ was also studied extensively in the context of functional thin films since its tetragonally stabilized form has a high dielectric constant. Müller[11] showed, that $ZrO_2$ behaves antiferroelectric: above a critical field in the order of 1 MV/cm, a temperature dependent field induced phase transformation to the f-phase occurs. A recent review of the current status of $HfO_2$, $ZrO_2$ and HZO based ferroelectric films can be found in[10].

Although the ferroelectricity in binary oxides is similar to perovskites, there are some differences which make these materials highly attractive for technological applications: the binary oxides do not suffer from a dead layer effect which makes perovskites ineffective for thin film technology. The mid range dielectric constant[8] of the binary oxides allows switching at moderate voltage, although the necessary field strength required for polarization reversal is much higher than for perovskites[14]. The large field strength without breakdown is possible due to the large band gap of 5.6 eV compared to 3-5 eV in perovskites[15].

Despite intense research in the last years to collect phenomenological knowledge and explore the fundamental effects, an open issue remains which is of some relevance for the fundamental understanding as well as for the technological applications: what is the exact reason for the stability of the ferroelectric phase in the various conditions where it has been found? As a fact, the bulk materials have not been stabilized in the ferroelectric f-phase so far, whereas the possible influencing factors on phase stability like temperature, doping and other defects, stress or strain, and surface or interface energy are known. It is under intensive investigation what specific set of circumstances is responsible for the existence of the experimental observed ferroelectric thin films.

The effect of doping to promote cubic Fm-3m (c-phase) or tetragonal $P4_2/nmc$ (t-phase) phases is well known from partially and fully stabilized ceramics. The ferroelectric f-phase was originally discovered in Si doped $HfO_2$[1]. So far all subsequently produced ferroelectric films were doped as well, with the exception of HZO films. However, to completely explain f-phase stability there is still a need for an additional mechanism since the ferroelectric properties typically disappear in thick films and bulk materials[16]. In this paper, we want to focus on the binary mixed oxide $Hf_{0.5}Zr_{0.5}O_2$ to minimize computational load otherwise necessary for calculations with dopants.

Since the discovery of transformation toughening by Garvie[17], the importance of stress effects causing the transition of stabilized tetragonal into the monoclinic $P2_1/c$ phase (m-phase) is well known in $ZrO_2$. Asymmetric stress effects have also been discussed as an explanation for the appearance of ferroelectric $ZrO_2$ grains in cubic stabilized $ZrO_2$ [18]. Furthermore, stress effects in thin $Hf_{0.5}Zr_{0.5}O_2$ films have been used as a possible explanation of how a cap electrode affects the ferroelectric phase by Müller[11,19] as well as film thickness causing film stress variation and appearance of a ferroelectric phase by Park[16].

The effect of grain size on phase stability is well known from the work of Garvie[17] and has an important application in $ZrO_2$ thin films contained in the high-k dielectric of DRAM capacitors. These $ZrO_2$ films with grain size below 30 nm are typically in the t-phase and the involved surface energy relative to the m-phase was experimentally measured by Pitcher et al. and other groups[20,21] and calculated by Christensen and Carter[22]. The same effect is known for $HfO_2$ which undergoes a phase transformation at a critical particle size of about 4 nm, but the surface energy has only been measured for the m-phase[23]. Müller[2] observed the size effect in ferroelectric $Hf_{0.5}Zr_{0.5}O_2$ thin films by identifying the thinner 7.5 nm film as less ferroelectric and more antiferroelectric than the thicker 9 nm film. This investigation was deepened in the systematic work by Park[24] who found a loss of ferroelectricity for films above 20 nm.

In this paper, we investigate the stability of the f-phase with density functional calculations (DFT). Sec. II. describes the computational setup. In Sec. III A-C the computed structures, total and free energies for various polymorphs are presented and compared with literature data. The phonon modes and the infrared (IR) spectra are computed as well. Since absolute values for total energies in DFT calculations are known to depend on the exchange-correlation functional and pseudopotential on the level of a few meV/f.u. (f.u.=formula unit $MO_2$, M=Hf, Zr), we validate our choice by comparison with available data. In Subsec. D the effects of stress and strain together with elastic properties are included, and possible scenarios for a stress or strain stabilization are discussed. In Subsec. E the effects of surface energy are included in form of a model. The model contains free parameters in the form of surface energies based on the limited amount of experimental and calculated data available so far. The results are discussed in light of a variety of experimental results. Subsec. F contains the computed effect of an electric field on the phase stability as well as the dielectric and ferroelectric properties. The conclusions are presented in Sec. IV.

## II. COMPUTATIONAL METHODS

All calculations are performed using the ABINIT package[25,26]. The exchange-correlation energy is computed with a local-density approximation using the Perdew-Wang parameterization. For Hf we use a self made norm-conserving Troullier-Martin pseudopotential generated with the fhi98 generator[27] to achieve good values for the lattice constants. For Zr and O we use well calibrated RRKJ norm-conserving pseudopotentials from the Opium project [31]. The levels Hf(5s,5p,5d,6s), Zr(4s,4p,4d,5s), and O(2s,2p) are treated as valence states with the following atomic valence configurations for the reference state: Hf($5s^2 5p^6 5d^2 6s^2$), Zr($4s^2 4p^6 4d^0 5s^0$), and O($2s^2 2p^4$). For Hf, core radii of 1.3 $a_0$, 1.8 $a_0$, 2.7 $a_0$ (Bohr radii) are chosen to describe angular waves from s to d. For Zr the cutoff radii were 1.58 $a_0$, 1.73 $a_0$ and 1.79 $a_0$ whereas for the O pseudopotential, a cutoff radius of 1.50 $a_0$ for both s and p waves is applied. We also adopted a separable form for the pseudopotentials treating the following angular momentum waves as local: s for Hf, f for Zr, and p for O. For Hf, a nonlinear core correction of radius 1.05 $a_0$ was used[28].

The Brillouin zone is sampled using the Monkhorst-Pack [29] scheme with a 6x6x6 k-point grid for all 12 atomic phases and a 3x6x6 for the 24 atomic orthorhombic Pbca phase (o-phase). The cutoff energy $E_{cut}$ of the plane-wave expansion was fixed at 30 Ha. For the calculation of the phonon modes a 2x2x2 q-point grid was calculated and interpolated to a 30x30x30 grid [30]. Integration of the phonon density of states yields the free energy. All parameters were carefully tested for convergence. The acoustic sum rule was imposed.

HZO cells are constructed by replacing 50 % of Zr atoms in $ZrO_2$ with Hf. There are two inequivalent possibilities to build such a cell, but the difference in total energy and other properties between both is negligible small. To save CPU time we choose one (see SI[32], TABLE SI) for our calculations.

Berry phase calculations were performed under an incrementally increasing electric field to study the effect of an external electrical field on m-, t- and f-phase cells. Structural relaxations are performed after each increase of 1 MV/cm in electric field strength.

## III. RESULTS

### A. Structural Properties

$HfO_2$ and $ZrO_2$ are structurally and chemically similar [33] and can adopt a variety of crystalline phases. Increasing the temperature, the monoclinic phase (m-phase, No. 14, space group (SG): $P2_1/c$) transforms (between 1270 K and 1370 K for $ZrO_2$[3] and at about 2073 K for $HfO_2$ [3,34]) into a tetragonal phase (t-phase, No. 137, SG: $P4_2/nmc$) and further (at about 2650 K for $ZrO_2$ [3] and about 2900 K for $HfO_2$[3]) into a cubic phase (c-phase, No. 225, SG: Fm3m). In addition, various orthorhombic phases exist: an orthorhombic I phase (o-phase, No. 61, SG: Pbca), an orthorhombic II phase (oII-phase, No. 62, SG: Pnma) and a polar orthorhombic III phase (f-phase, No. 29, SG: $Pca2_1$). The phase transformation from the m- to the o-phase can be observed at a compressive pressure of about 4-12 GPa in bulk for $HfO_2$[35,36] and $ZrO_2$[37,38]. The oII-phase occurs at a very high compressive pressure above 20 GPa for both $HfO_2$[36] and $ZrO_2$[38]. We do not include this high pressure phase in our study, since such high pressures are irrelevant for most thin film applications. The transformation to the polar f-phase can be observed under asymmetric stress conditions[18] which will be discussed in section C. Although the unit cell of the c-phase has only 3 and the t-phase only 6 atoms, we represent all the structures except the o-phase in 12 atom cells for a better visualization of the structural similarity. The Wyckoff coordinates are shown in TABLE I and the structures in FIG. 1[39]. The polarization of the f-phase can be visualized in the form of the displacement of the four O1 atoms in the z-direction after polarization reversal by mirror reflection, whereas the metal and O2 atoms are hardly displaced. When the motion of an atom $i$ in direction $j$ is $\Delta r_{ij}$ relative to the mirror symmetric, polarization neutral configuration, the polarization $P_j$ of the structure is

$$P_j = \frac{e}{V} \cdot \sum_{i=1}^{12} Z_{ij}^* \cdot \Delta r_{ij} \qquad (1)$$

with cell volume $V$, Born charge tensor $Z^*_{ij}$ and unit charge $e$. Upon a point reflection the polarization changes as well, but the resulting structure is unequal to the mirror reflected structure. The reason is that the f-phase appears to occur in two different forms with opposite chirality. We are unaware of any physical reason to discriminate in favor of any particular chiral form as both structures have the same total energies and hence decided for the mirror reflected structure shown in FIG. 1 based upon better visualization of the polarization reversal. Furthermore, the 24 atomic unit cell of the o-phase can be constructed by gluing the two illustrated unit cells of the f-phase with opposite polarization and chirality alongside the xy-plane together and relaxing the structure. Therefore the o-phase has one additional symmetry compared to the f-phase.

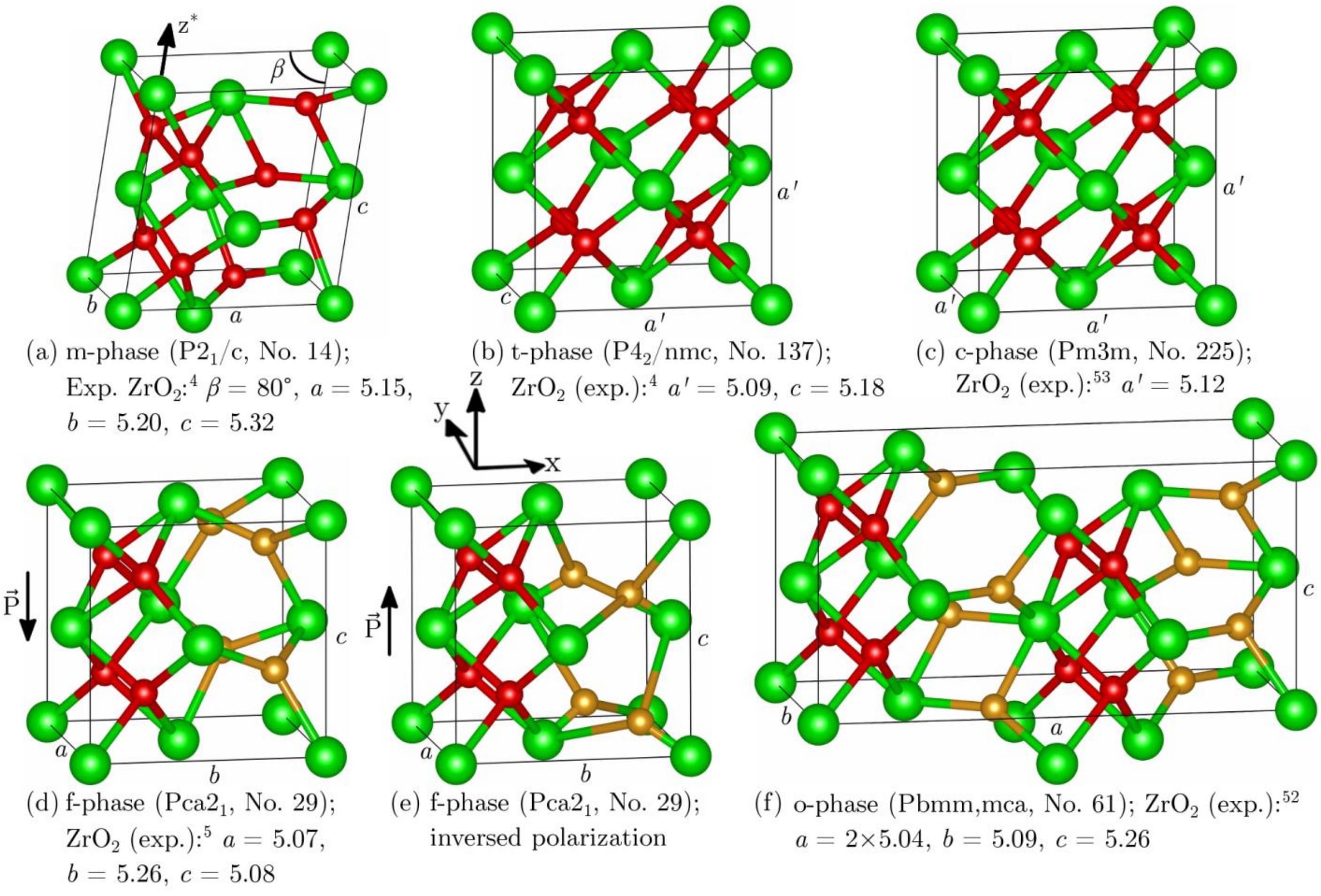

FIG. 1. Crystal phases[39] and experimental lattice constants in Å for $ZrO_2$. Zr atoms are represented in green (big), O atoms in red (small). In the f-phases the O atoms mainly responsible for the polarization are highlighted golden. To illustrate the relationship between the f- and o-phase, the corresponding O atoms are highlighted in the o-phase as well. The polarization axis of the two f-phase cells is marked by black arrows (P).

The similarity between structures can be better understood by comparing the relative coordinates of the Wyckoff positions in an appropriate representation (choice of origin). The chosen representation allows the construction of an initial reaction path for the calculation of the phase transformation to determine the transition state. The relative coordinates and cell parameters of the calculated structures are shown in TABLE I. The representations of the structures were checked with FINDSYM [40].

TABLE I. Calculated lattice constants $a$, $b$, and $c$ [41] for a unit cell in Å. $a'$, $b'$, and $c'$ are the lattice constants of a 12 atomic cell with less than 12 atoms per unit cell. The cell volume $V$ is in Å$^3$, and the total energy difference $\Delta U$ in meV/f.u. The lattice constants are aligned with the direction $x$, $y$, $z$, and $z^*$ (given in Miller vector notation) as defined in FIG. 1. $\beta$ is the angle of the m-phase between $x$ and $z^*$. The Wyckoff positions of the Metals (M) and Oxygen atoms (O1 and O2) are given in relative coordinates. The arrows indicate the polarization direction of the f-phase.

| | | | | | | | |
|---|---|---|---|---|---|---|---|
| | | c-$ZrO_2$ | | | c-$HfO_2$ | | |
| | | $\Delta U=96$ | $V=129.2$ | | $\Delta U=112$ | $V=128.4$ | |
| | | $x=[100]$ | $y=[010]$ | $z=[001]$ | $x=[100]$ | $y=[010]$ | $z=[001]$ |
| | | $a'=5.05$ | $b'=5.05$ | $c'=5.05$ | $a'=5.04$ | $b'=5.04$ | $c'=5.04$ |
| M | b | 0.250 | 0.000 | 0.250 | 0.250 | 0.000 | 0.250 |
| O1 | c | 0.000 | 0.250 | 0.000 | 0.000 | 0.250 | 0.000 |
| O2 | c | 0.500 | 0.250 | 0.500 | 0.500 | 0.250 | 0.500 |
| | | t-$ZrO_2$ | | | t-$HfO_2$ | | |
| | | $\Delta U=49$ | $V=131.8$ | | $\Delta U=92$ | $V=131.1$ | |
| | | $x=[110]$ | $y=[001]$ | $z=[\underline{1}10]$ | $x=[110]$ | $y=[001]$ | $z=[\underline{1}10]$ |
| | | $a'=5.06$ | $c=5.15$ | $b'=5.06$ | $a'=5.05$ | $c=5.14$ | $b'=5.05$ |
| M | b | 0.250 | 0.000 | 0.250 | 0.250 | 0.000 | 0.250 |
| O1 | d | 0.000 | 0.294 | 0.000 | 0.000 | 0.293 | 0.000 |
| O2 | d | 0.500 | 0.206 | 0.500 | 0.500 | 0.207 | 0.500 |
| | | m-$ZrO_2$ | | | m-$HfO_2$ | | |
| | | $\Delta U=0$ | $V=138.5$ | $\beta=80.5°$ | $\Delta U=0$ | $V=138.4$ | $\beta=80.3°$ |
| | | $x=[100]$ | $y=[010]$ | $z^*=[001]$ | $x=[100]$ | $y=[010]$ | $z^*=[001]$ |
| | | $a=5.11$ | $b=5.22$ | $c=5.27$ | $a=5.12$ | $b=5.18$ | $c=5.29$ |
| M | 4e | 0.278 | 0.043 | 0.290 | 0.276 | 0.042 | 0.292 |
| O1 | 4e | 0.074 | 0.334 | 0.165 | 0.068 | 0.331 | 0.153 |
| O2 | 4e | 0.553 | 0.258 | 0.481 | 0.550 | 0.257 | 0.480 |
| | | f-$ZrO_2$ ↓ | | | f-$HfO_2$ ↓ | | |
| | | $\Delta U=37$ | $V=133.8$ | | $\Delta U=62$ | $V=133.5$ | |
| | | $x=[0\underline{1}0]$ | $y=[100]$ | $z=[001]$ | $x=[0\underline{1}0]$ | $y=[100]$ | $z=[001]$ |
| | | $b=5.04$ | $a=5.25$ | $c=5.06$ | $b=5.04$ | $a=5.23$ | $c=5.06$ |
| M | 4a | 0.268 | 0.032 | $\frac{1}{4}+0.002$ | 0.267 | 0.032 | $\frac{1}{4}+0.007$ |
| O1 | 4a | 0.073 | 0.369 | $\frac{1}{4}-0.133$ | 0.068 | 0.389 | $\frac{1}{4}-0.138$ |
| O2 | 4a | 0.539 | 0.266 | $\frac{1}{2}+0.006$ | 0.537 | 0.267 | $\frac{1}{2}+0.008$ |
| | | f-$ZrO_2$ ↑ | | | f-$HfO_2$ ↑ | | |
| | | $\Delta U=37$ | $V=133.8$ | | $\Delta U=62$ | $V=133.5$ | |
| | | $x=[0\underline{1}0]$ | $y=[100]$ | $z=[001]$ | $x=[0\underline{1}0]$ | $y=[100]$ | $z=[001]$ |
| | | $b=5.04$ | $a=5.25$ | $c=5.06$ | $a=5.04$ | $b=5.23$ | $c=5.06$ |
| M | 4a | 0.268 | 0.032 | $\frac{1}{4}-0.002$ | 0.267 | 0.032 | $\frac{1}{4}-0.007$ |
| O1 | 4a | 0.073 | 0.369 | $\frac{1}{4}+0.133$ | 0.068 | 0.389 | $\frac{1}{4}+0.138$ |
| O2 | 4a | 0.539 | 0.266 | $\frac{1}{2}$-0.006 | 0.537 | 0.267 | $\frac{1}{2}$-0.008 |
| | | o-$ZrO_2$ | | | o-$HfO_2$ | | |
| | | $\Delta U$=26 $V$=138.7 | | | $\Delta U$=24 $V$=138.4 | | |
| | | $x$=[100] | $y$=[010] | $z$=[001] | $x$=[100] | $y$=[010] | $z$=[001] |
| | | $a/2$=5.07 | $b$=5.20 | $c$=5.27 | $a/2$=5.07 | $b$=5.15 | $c$=5.29 |
| M | 8c | 0.278/2 | 0.042 | 0.340 | 0.276/2 | 0.041 | 0.343 |
| O1 | 8c | 0.069/2 | 0.338 | 0.174 | 0.064/2 | 0.325 | 0.160 |
| O2 | 8c | 0.552/2 | 0.249 | 0.589 | 0.550/2 | 0.248 | 0.589 |

A martensitic phase transformation between crystal structures requires a specific spatial orientation due to constraints by transition path and barrier. The orientation between the t- and m-phase during temperature driven-phase transformations has been investigated experimentally and theoretically [42,43]. It was found that the $[010]_m$ || $[001]_t$ directions and the $(100)_m$ || $(110)_t$ planes coincide. In addition the spatial orientation between the t- and f-phase has been investigated by Kisi [4] who found a similar coincidence of $[010]_f$ || $[001]_t$ and $(100)_f$ || $(110)_t$. The spatial orientation between the f- and o-phase is derived from the representation in FIG. 1 and implies a spatial relation between the o- and m-phase via the f- and t-phase.

In the following discussion, we will focus on $ZrO_2$ to further exemplify the relation between the various phases (Values from TABLE I). The m→t transition involves a volume change from 138.5 Å$^3$ to 131.8 Å$^3$ and the change of $c$ from 5.27 Å to 5.06 Å. Besides the difference of the lattice angle $\beta$, the main distinction between the m- and the f-phase is a polarization dependent displacement of the O1 atom in z-direction. The o-phase is closely related to the f-phase because the 24 atomic cell is composed of two oppositely polarized 12 atomic f-phase cells. Nonetheless the cell volume and lattice constants differ: the volume changes from 138.7

Å$^3$ to 133.8 Å$^3$ and the length of the polarization axis $c$ from 5.27 Å to 5.06 Å. The experimental results show only a small difference in volume and length between these two phases. A transformation from the m-phase to the f-phase involves a volume change from 138.5 Å$^3$ to 133.8 Å$^3$ and a reduced length of the polarization direction $c$ from 5.27Å to 5.06Å. The transformation from the t-phase to the f-phase involves only a volume change from 131.8 Å$^3$ to 133.8 Å$^3$ and the polarization direction $c$ has almost no change. At this point, a transformation to the f-phase seems to be easiest from the t-phase although both phases have no close symmetry relationship. Comparing Hf and Zr, all the $HfO_2$ cells are slightly smaller than the $ZrO_2$ cells consistent with the lanthanide contraction argument [33].

It is known that the accuracy of the calculated results depend on the chosen density functional and pseudo-potential. While lattice constants from generalized gradient approximation (GGA) calculations are typically too large, the results from local density approximation (LDA) calculations are typically too small (about 1 %). However, the discrepancy for the LDA decreases, when the expansion effect of zero point motion is included [44]. Furthermore, the calculated results depend on the chosen basis set and the number of k-points which were not always well documented in previous studies. In TABLE II, III, and IV we compare our calculated results for lattice constants of $HfO_2$, HZO, and $ZrO_2$ respectively, with calculated values and with experimental data found in literature about the investigated phases. For HZO, only our own calculated values are available as well as experimental data from Müller et al. [11]. In all cases, except for the c-phases, our structural values have proved to be better than 1 % compared to experimental data without taking expansion effects from zero point motions into account.

TABLE II. Comparison lattice constants $a$, $b$, and $c$ in Å and total energy difference $\Delta U$ in meV/f.u. for $HfO_2$

| Structure | $\Delta U^a$ [meV] | $V$ [Å$^3$] | $A_{xz}$ [Å$^2$] | $a$ [Å] | $b$ [Å] | $c$ [Å] | method reference |
|---|---|---|---|---|---|---|---|
| m-$HfO_2$ | / | 137.9 | 26.7 | 5.12 | 5.17 | 5.29 | exp.[45] |
| | / | 135.8 | 26.4 | 5.07 | 5.14 | 5.29 | exp.[11] |
| | 0 | 137.1 | 26.6 | 5.11 | 5.16 | 5.28 | LDA [(*)] |
| | 0 | 136.1 | 26.4 | 5.09 | 5.16 | 5.26 | LDA[46] |
| | 0 | 125.3 | 24.8 | 4.95 | 5.06 | 5.08 | LDA[47] |
| | 0 | 130.9 | - | - | - | - | LDA[48] |
| | 0 | 139.0 | 26.8 | 5.13 | 5.19 | 5.30 | GGA[48] |
| | 0 | 139.8 | 26.9 | 5.14 | 5.20 | 5.31 | GGA[49] |
| | 0 | 138.2 | 26.6 | 5.12 | 5.19 | 5.28 | PAW[46] |
| | $\Delta U^a$ | $V$ | $A_{xz}$ | $a'$ | $c'$ | $b'$ | |
| t-$HfO_2$[b] | / | 133.1 | 25.6 | 5.06 | 5.20 | - | exp.[45] |
| | / | 133.1 | 25.6 | 5.06 | 5.20 | - | exp.[50] |
| | 92 | 129.5 | 25.3 | 5.03 | 5.12 | - | LDA [(*)] |
| | 71 | 118.8 | 24.0 | 4.90 | 4.95 | - | LDA[47] |
| | >60 | 135.2 | 25.6 | 5.06 | 5.28 | - | LDA[49] |
| | 99 | 125.1 | - | - | - | - | LDA[48] |
| | 156 | 133.1 | 25.6 | 5.06 | 5.20 | - | GGA[48] |
| | 138 | 130.3 | 25.3 | 5.03 | 5.15 | - | PAW[46] |
| | $\Delta U^a$ | $V$ | $A_{xz}$ | $b$ | $a$ | $c$ | |
| f-$HfO_2$ | / | - | - | - | - | - | exp. |
| | 62 | 132.1 | 25.3 | 5.02 | 5.22 | 5.04 | LDA [(*)] |
| | 48 | 121.0 | 23.9 | 4.88 | 5.07 | 4.89 | LDA[47] |
| | 24 | 138.1 | 26.1 | 5.10 | 5.30 | 5.11 | LDA[6] |
| | >30 | 134.6 | 25.5 | 5.01 | 5.29 | 5.08 | GGA[49] |
| | 63 | 102.8 | 24.1 | 4.90 | 5.11 | 4.92 | PAW[46] |
| | $\Delta U^a$ | $V$ | $A_{xz}$ | $a/2$ | $b$ | $c$ | |
| o-$HfO_2$[c] | / | 132.6 | 26.2 | 5.01 | 5.06 | 5.23 | exp.[51] |
| | 24 | 138.4 | 26.7 | 5.07 | 5.15 | 5.29 | LDA [(*)] |
| | 60 | 137.8 | 27.0 | 5.08 | 5.11 | 5.31 | LDA[6] |
| | 29 | 126.0 | - | - | - | - | LDA[48] |
| | 65 | 134.1 | 25.4 | 4.92 | 4.96 | 5.16 | GGA[48] |
| | >25 | 134.4 | 26.5 | 5.02 | 5.08 | 5.27 | GGA[49] |
| | $\Delta U^a$ | $V$ | $A_{xz}$ | $a'$ | $b'$ | $c'$ | |
| c-$HfO_2$[b] | / | 131.1 | 25.8 | 5.08 | - | - | exp.[52] |
| | 137 | 127.3 | 25.3 | 5.03 | - | - | LDA [(*)] |
| | 93 | 116.9 | 23.9 | 4.89 | - | - | LDA[47] |
| | 152 | 123.0 | - | - | - | - | LDA[48] |
| | 237 | 129.6 | 25.6 | 5.06 | - | - | GGA[48] |

| | 208 | 127.3 | 25.3 | 5.03 | - | - | PAW[46] |
|---|---|---|---|---|---|---|---|

(*) this work, [a]relative to m-$HfO_2$, [b]values for 12 atom cell, [c]half c-axis

TABLE III Comparison lattice constants *a*, *b*, and *c* in Å and total energy difference $\Delta U$ in meV/f.u.for HZO.

| Structure | $\Delta U^{a}$ [meV] | $V$ [Å$^3$] | $A_{xz}$ [Å$^2$] | $a$ [Å] | $b$ [Å] | $c$ [Å] | Method reference |
|---|---|---|---|---|---|---|---|
| m-HZO | 0 | 137.6 | 26.6 | 5.11 | 5.18 | 5.28 | LDA (*) |
| | $\Delta U^{a}$ | $V$ | $A_{xz}$ | $a'$ | $c$ | $b'$ | |
| t-HZO[b] | 70 | 130.3 | 25.4 | 5.04 | 5.13 | =a | LDA (*) |
| | $\Delta U^{a}$ | $V$ | $A_{xz}$ | $b$ | $a$ | $c$ | |
| f-HZO | / | 132.3 | 25.3 | 5.01 | 5.24 | 5.04 | exp.[11] |
| | 49 | 132.8 | 25.4 | 5.03 | 5.23 | 5.05 | LDA (*) |
| | $\Delta U^{a}$ | $V$ | $A_{xz}$ | $a/2$ | $b$ | $c$ | |
| o-HZO[c] | 25 | 137.1 | 26.7 | 5.06 | 5.14 | 5.27 | LDA (*) |
| | $\Delta U^{a}$ | $V$ | $A_{xz}$ | $a'$ | $b'$ | $c'$ | |
| c-HZO[b] | 119 | 127.3 | 25.3 | 5.03 | =a | =a | LDA (*) |

(*) this work, [a]relative to m-HZO, [b]values for 12 atom cell, [c]half c-axis

TABLE IV. Comparison lattice constants *a*, *b*, and *c* in Å and total energy difference $\Delta U$ in meV/f.u.for $ZrO_2$.

| Structure | $\Delta U^{a}$ [meV] | $V$ [Å$^3$] | $A_{xz}$ [Å$^2$] | $a$ [Å] | $b$ [Å] | $c$ [Å] | Method reference |
|---|---|---|---|---|---|---|---|
| m-$ZrO_2$ | / | 140.3 | 27.0 | 5.15 | 5.20 | 5.32 | exp.[4] |
| | 0 | 138.2 | 26.6 | 5.11 | 5.20 | 5.28 | LDA (*) |
| | 0 | 136.6 | 26.3 | 5.09 | 5.20 | 5.24 | LDA[47] |
| | 0 | 136.7 | - | - | - | - | LDA[48] |
| | 0 | 136.1 | 26.3 | 5.09 | 5.18 | 5.24 | LDA[53] |
| | 0 | 144.7 | 27.4 | 5.20 | 5.28 | 5.35 | GGA[48] |
| | 0 | 144.1 | 27.5 | 5.19 | 5.24 | 5.38 | GGA[53] |
| | $\Delta U^{a}$ | $V$ | $A_{xz}$ | $a'$ | $c$ | $b'$ | |
| t-$ZrO_2$[b] | / | 134.2 | 25.9 | 5.09 | 5.18 | =a | exp.[4] |
| | / | 133.4 | 25.8 | 5.08 | 5.17 | =a | exp.[11] |
| | 49 | 131.3 | 25.5 | 5.05 | 5.15 | =a | LDA (*) |
| | 34 | 130.1 | 25.4 | 5.04 | 5.12 | =a | LDA[47] |
| | 38 | 130.4 | - | - | - | - | LDA[48] |
| | 50 | 129.3 | 25.3 | 5.03 | 5.11 | =a | LDA[53] |
| | 109 | 137.9 | 26.3 | 5.13 | 5.21 | =a | GGA[48] |
| | 112 | 138.4 | 26.2 | 5.12 | 5.28 | =a | GGA[53] |
| | $\Delta U^{a}$ | $V$ | $A_{xz}$ | $b$ | $a$ | $c$ | |
| f-$ZrO_2$ | / | 135.5 | 25.8 | 5.07 | 5.26 | 5.08 | exp.[5] |
| | 37 | 133.4 | 25.5 | 5.04 | 5.24 | 5.05 | LDA (*) |
| | 34 | 132.1 | 25.3 | 5.02 | 5.22 | 5.04 | LDA[47] |
| | -60 | 138.1 | 26.1 | 5.10 | 5.30 | 5.11 | LDA[6] |
| | $\Delta U^{a}$ | $V$ | $A_{xz}$ | $a/2$ | $b$ | $c$ | |
| o-$ZrO_2$[c] | / | 134.9 | 26.5 | 5.04 | 5.09 | 5.26 | exp.[54] |
| | 25 | 138.7 | 26.8 | 5.07 | 5.20 | 5.27 | LDA (*) |
| | -99 | 137.8 | 27.0 | 5.08 | 5.11 | 5.31 | LDA[6] |
| | 14 | 131.9 | - | - | - | - | LDA[48] |
| | 26 | 130.8 | 26.0 | 4.99 | 5.03 | 5.21 | LDA[53] |
| | 49 | 138.8 | 26.1 | 5.09 | 5.14 | 5.31 | GGA[48] |
| | 67 | 138.1 | 26.9 | 5.08 | 5.13 | 5.30 | GGA[53] |
| | $\Delta U^{a}$ | $V$ | $A_{xz}$ | $a'$ | $b'$ | $c'$ | |
| c-$ZrO_2$[b] | / | 134.2 | 26.2 | 5.12 | - | - | exp.[55] |
| | 120 | 127.3 | 25.3 | 5.03 | - | - | LDA (*) |
| | 82 | 127.3 | 25.3 | 5.03 | - | - | LDA[47] |
| | 67 | 128.6 | ? | ? | - | - | LDA[48] |
| | 94 | 127.3 | 25.3 | 5.03 | - | - | LDA[53] |
| | 171 | 134.8 | 26.3 | 5.13 | - | - | GGA[48] |
| | 215 | 134.9 | 26.2 | 5.12 | - | - | GGA[53] |

(*) this work, [a]relative to m-$ZrO_2$, [b]values for 12 atom cell, [c]half c-axis

For the calculation of the force required for a stress induced phase transformation, it is important that the spatial orientations of the cells match together for the phase transformation, and the calculated cell parameters and volumes agree with experimental data at least in their relative values to each other. Otherwise values for the necessary pressure or stress will be inaccurate. For the plane strain, the cell areas in the compressed or expanded planes are relevant. The comparison of our calculated volume ratios to the experimental volume ratios are summarized in the supplementary material (SI[32], TABLE SII) shows good agreement.

B. Total Energy

The structural relaxation provides values for the specific total energy per f.u. for each phase. TABLE I contains the calculated total energy differences $\Delta U$ relative to the m-phase. These values constitute the foundation of the calculation of the phase stability since the relevant criteria like the Helmholtz free energy and the Gibbs energy are calculated from this difference by adding a temperature or pressure contribution. The total energy values are only indirectly accessible in experiments and their consistency can only be concluded from the correctness of the final result. LDA results repeatedly in smaller energy differences than GGA calculations. These differences have been systematically studied by Jaffe et al. [48] for both $HfO_2$ and $ZrO_2$, and by Fadda et al. [53] for $ZrO_2$ both using ultrasoft pseudopotentials. Considering the total energy, Jaffe et al. concluded that GGA results are needed since with LDA the phase transformation pressure between the m- and o-phase turned out to be negative. Fadda et al.

compared the calculated results with plenty of data sources and found LDA as a better overall description but found too small values for the transition pressure between m- and o-phase as well. A comparison of their results for *ΔU* in TABLE IV shows that Jaffe et al. obtained systematically smaller values for all phases and compounds than Fadda et al. Also, their with LDA calculated volume ratios of the structures are not in good agreement with experimental results. Therefore the comparison shows the dependence of the correctness of the conclusion on the quality of the pseudopotentials.

The results for *ΔU* show consistently positive values for all authors for the stability of f-$HfO_2$. However, Lowther[6] revealed large negative values for f-$ZrO_2$ and even larger negative values for the o- $ZrO_2$, implying stability of o-phase. With this single exception, all calculated results lead to same order of increasing total energy values: $U(m)<U(o)<U(f)<U(t)<U(c)$.

C. Helmholtz Free Energy

To assess the phase stability at a finite temperature $T$ and entropy $S$, we calculated the Helmholtz free energy $F$ as

$$F = U - TS \tag{2}$$

from the phonon contribution by integrating over the phonon density of state. Fadda [53] and Zhao [56] have argued, that LDA gives better results for phonon mode frequencies than GGA. The quality of our LDA and the choice of pseudopotentials can be estimated by comparison of the calculated IR-mode frequencies with an experiential IR-absorption spectrum (see FIG. 2 a-c). With a slight underestimation of the computed frequencies the agreement with experimental data is generally good. The similarity of the structures of $Hf_{1-\chi}Zr_{\chi}O_2$ becomes visible in the IR-absorption spectrum for $\chi = 0$, 0.5 and 1 (see SI [32], FIG. S1).

The IR-absorption spectra as well as the Raman spectra are fingerprints to identify different phases. The difficulty to distinguish the orthorhombic f- and o-phase from the XRD spectrum has been discussed by Howard [57]. In contrast, according to the ab initio results the t-, f- and the o-phase might be more easily distinguished using IR- or Raman-spectra instead of XRD (see SI [32], FIG. S1 a-c).

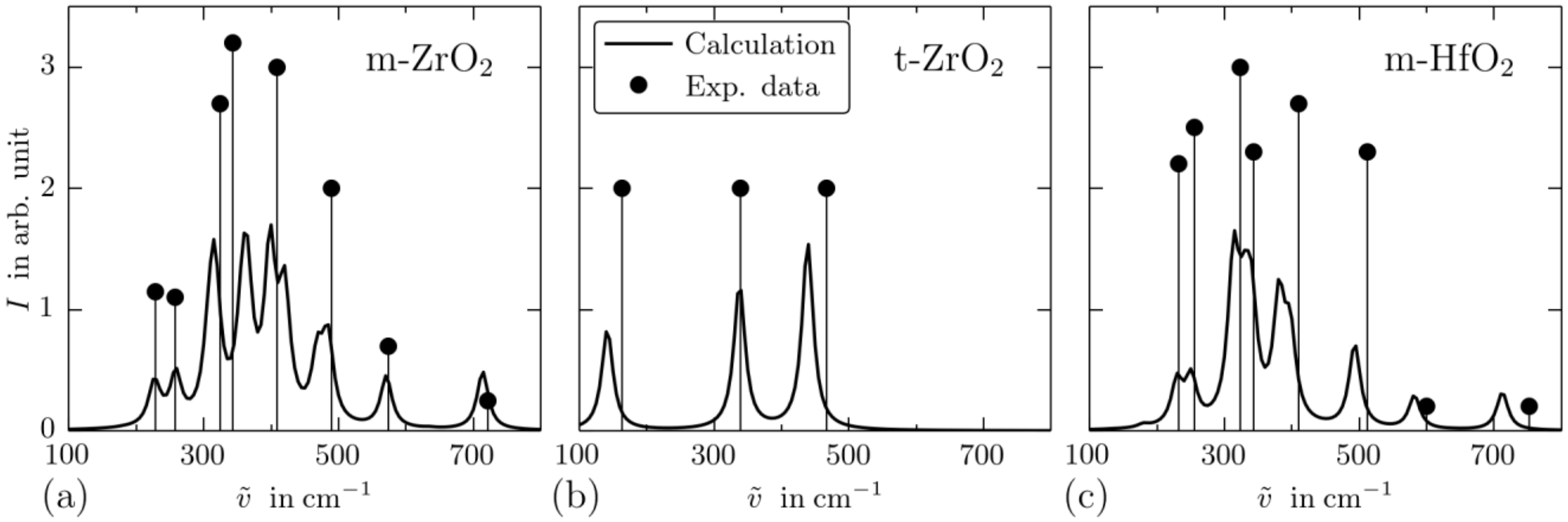

FIG. 2. (a) Calculated IR-Spectrum of (a) $m\text{-}ZrO_2$, (b) $t\text{-}ZrO_2$ and (c) $m\text{-}HfO_2$ in comparison with experimental [58] [59].

The results for the Helmholtz free energy difference *ΔF(T)* of the phases relative to the m-phase as a function of temperature are shown in FIG. 3 a-c for $HfO_2$, HZO, and $ZrO_2$. Qualitatively similar results for $HfO_2$, with the exception of the c-phase, have been obtained by Huan [49], and for $ZrO_2$ by Fadda et al. [53] with the exception of the f- and c-phase. For all values $\chi$ in $Hf_{1-\chi}Zr_{\chi}O_2$ the m-phase has the lowest Helmholtz free energy until the m→t transition temperature (2100 K, 1750 K and 1250 K for $HfO_2$, HZO and $ZrO_2$ respectively compared to 2073 K, 1700 K and 1300 K in experiments [3,34]). At a slightly higher temperature (the t→c transition temperature) the c-phase acquires the lowest Helmholtz free energy (2900 K, 2775 K and 2650 K in experiment [3] compared to 1950 K, 1550 and 1100 K in calculation for $HfO_2$, HZO and $ZrO_2$ respectively). Our transition temperatures for the m→t transition are in good agreement with experimental data. However, the order of the m→t and t→c transformations is reversed in our calculation, indicating either a deficiency of the chosen pseudopotentials for the c-phase or a limited validity of the harmonic approximation for large temperatures. All Helmholtz free energy differences are larger in $HfO_2$ by about 20 % compared to $ZrO_2$.

More important for our paper is the result, that neither the f- nor the o-phase becomes favored by a increase in temperature, although the f-phase is only about 50 meV/f.u. and the o-phase only about 25 meV/f.u. less favorable than the m-phase. This is consistent with the results of Huan[49] for $HfO_2$ and can be understood by an entropy argument: the t- and c-phase lose their higher symmetry faster with increasing thermal motion than the m-, o-, and f-phase and are therefore involved in thermally driven phase transformations. The degree of symmetry of the m-, o-, and f-phase is the same, measured by the number of symmetry operations per atom (=0.333), explaining their nearly temperature independent Helmholtz free energy difference. Therefore the o-phase can not be favored in a thermally driven phase transformation.

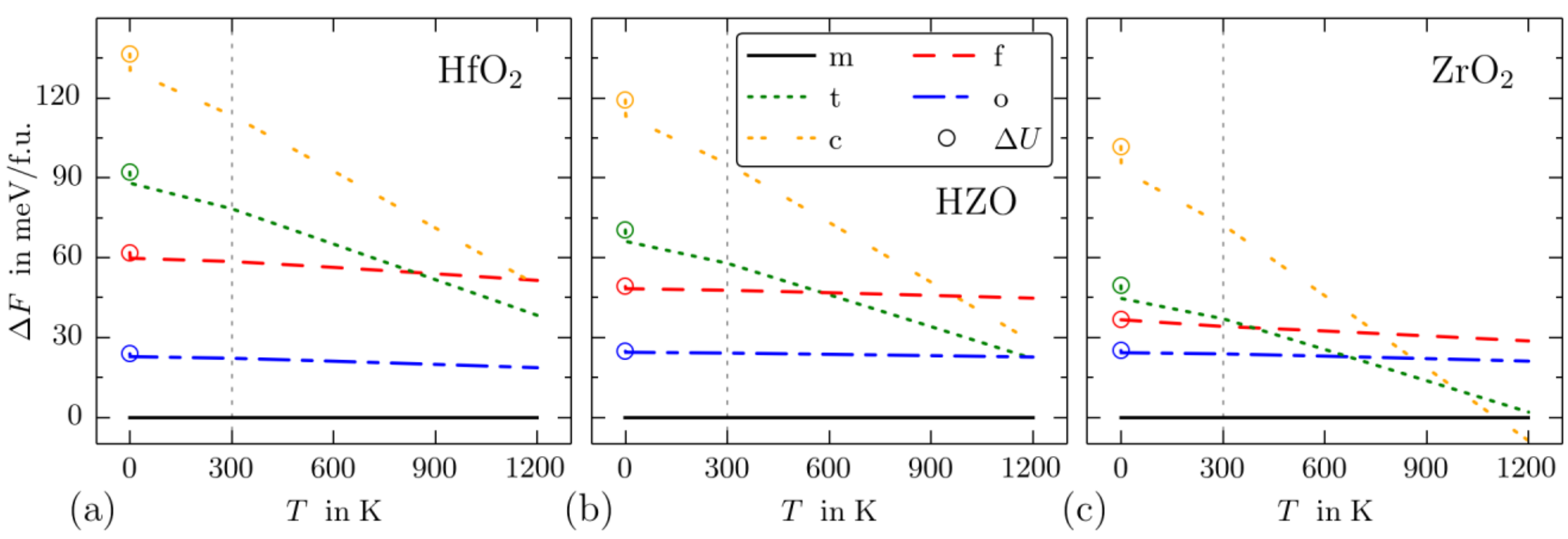

FIG. 3. Calculated total energy differences *ΔU* (o) and Helmholtz free energy differences *ΔF(T)* relative to the m-phase as a function of temperature, for (a) $HfO_2$ (b) HZO and (c) $ZrO_2$.

### D. Strain and Stress Gibbs Energy

Since it has been established that temperature does not stabilize the f-phase alone, we now investigate stress effects for a possible stabilization mechanism. Using an isotropic pressure constraint the phase stability is determined by the Gibbs energy $G$ as

$$G = F + pV = U - TS + pV \tag{3}$$

with the pressure $p$ and the volume $V$. Diamond-anvil cells have been used to determine pressure driven phase transformations experimentally. For $HfO_2$[36,60] and $ZrO_2$[38,54], the phase transformation from the m-phase to an orthorhombic phase has been found between 4-12 GPa. After a long dispute, the phase was identified as the o-phase in both cases [4,37]. To obtain a good value of the m→o transition pressure in the simulation the total energy difference between the o- and m-phase, and the value of the bulk modulus must be calculated accurately enough ([48,53] and discussion in sect. B). Our calculated values for transition pressure and bulk modulus compared to experimental data and other calculated values from the literature can be found in SI [32], TABLE SIII. With values between 10-15 GPa for the m→o phase transformation we obtain a good agreement. The phase transformation pressure has been derived from values of the Gibbs energy difference $\Delta G$ at 300 K relative to the m-phase in FIG. 4(a). The m-phase is stable around zero pressure. For compressive pressure above 10-15 GPa the o-phase is favored. The t-phase becomes stable for tensile pressures, but the f-phase is never stabilized. The Gibbs energy and hence the transition pressure is modified by effects of the surface energy as known from diamond-anvil experiments with 30 nm small grain material from Ohtaka[54]. The model for the inclusion of the surface energy to the Gibbs / Helmholtz free energy will be discussed in Subsection E. FIG. 4(b) illustrates calculated values for a grain size of 30nm. The surface effects lower the phase transformation pressure.

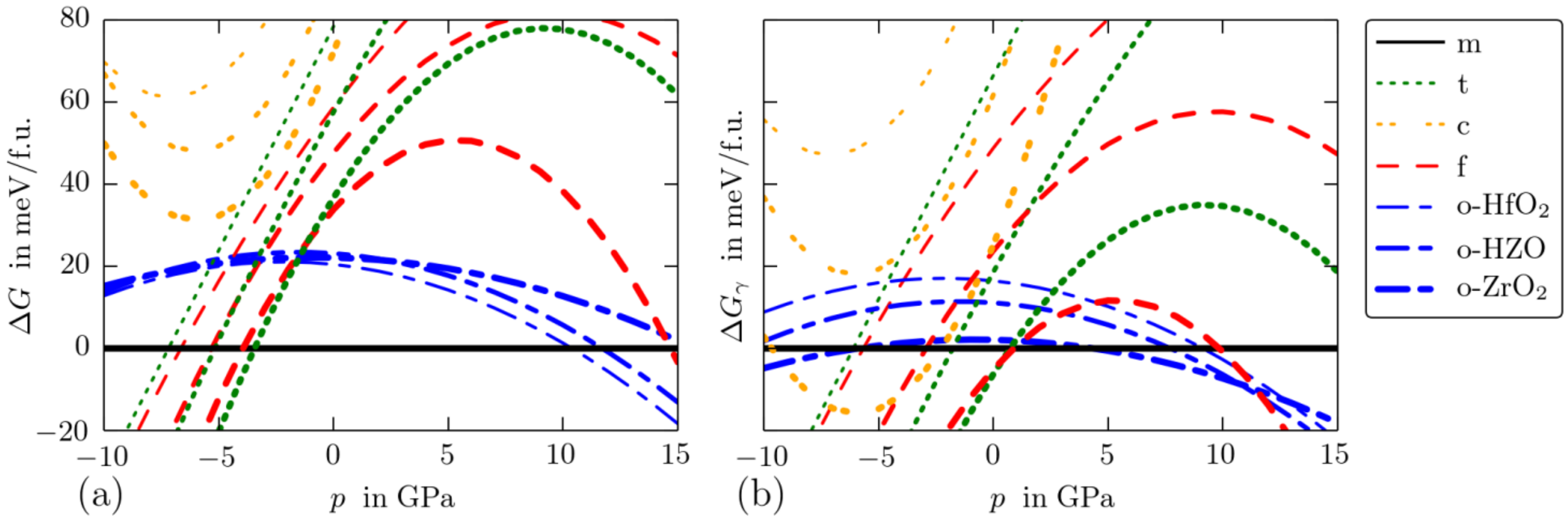

FIG. 4. (a) Calculated Gibbs energy difference $\Delta G$ relative to the m-phase around zero pressure and m→o transition for compressive pressure. (b) Gibbs energy difference including the surface energy $\Delta G_\gamma$ for 30nm grains according to our model.

The fact that the f-phase has only been observed in thin films suggests that the phase could be stabilized under anisotropic strain conditions determined from a fixed surface area and zero stress in the normal direction of the film. The stable phase under any strain constraint within the three strain planes $xy$, $xz$, and $yz$, is the one with the lowest value of the Helmholtz free energy[61]. In principle, all combinations of strain planes have to be considered while comparing the strain energy of two phases. However, an energy crossover can only happen in spatial orientations of cells with a rough match of lattice constants. For our calculations of the energy crossover under strain we have chosen the spatial orientations of the 12 atomic cells as shown in TABLE I. The spatial orientation between the c- and t-phase is obvious, as is the spatial orientation between o- and f-phase. For the t-phase, the c-axis in the crystallographic [001] direction is the largest with 5.15 Å and conforms most easily to the a-axis in [100] direction of the f-phase of length 5.25 Å. The *a*- and *b*-directions of length 5.06 Å conform to the *b*- and *c*-directions of the f-phase of length 5.04 Å and 5.06 Å. The remaining two possibilities of a match: transformation ($a^{(t)}$, $c^{(t)}$, $b^{(t)}$)→($b^{(f)}$, $a^{(f)}$, $c^{(f)}$) (choice in TABLE I) and $c^{(f)}$ and $b^{(f)}$ inverted are obviously equivalent. Therefore a t→f transformation requires only a moderate elongation of the $c^{(t)}$-axis and no significant modification of the $a^{(t)}$ and $b^{(t)}$ axis and could be promoted with a strain in the *xz*- or *yz*-plane. Crucial is the strain energy of the m- and o-phase since these start with the lowest energy from the beginning. The relative orientation between the m- and t-phase has been widely discussed in the literature in context of transformation toughening. Our choice for m→t in TABLE I corresponds to the findings of[42,43] ($a^{(m)}$, $b^{(m)}$, $c^{(m)}$)→( $a^{(t)}$, $c^{(t)}$, $b^{(t)}$). An alternative orientation ($a^{(m)}$, $b^{(m)}$, $c^{(m)}$)→($a^{(t)}$, $b^{(t)}$, $c^{(t)}$) was favored by Luo et al. [34] while he was looking for the smallest transition barrier between the m- and t-phase. We have calculated the total energy $\Delta U$ for the *xz*- and *yz*-strain planes, and all phases for the chosen orientation. The values of the lattice constants and the corresponding cell areas $A$ have been chosen by modifying the in plane stress. In equilibrium, the grains should then adopt the crystal phase with the minimum Helmholtz free energy. In FIG. 5 a-b we show the results as a function of cell area $A$ for $ZrO_2$. For $HfO_2$ and HZO see SI[32], FIG. S2. In our chosen orientation, only a small window of stability occurs for the f-phase in the *yz*-plane. The crossover point corresponds to a film stress of about 3 GPa in the f-phase and 8 GPa in the m-phase in the *yz*-plane (label m). No crossover exists in the *xz*-plane. In the orientation chosen by Luo et al.[34] there is a large window of stability in *xz*-plane, with corresponding plane stress of 9 GPa in the m-phase (label m') and near zero stress in f-phase. In the *yz*-plane both orientations coincide.

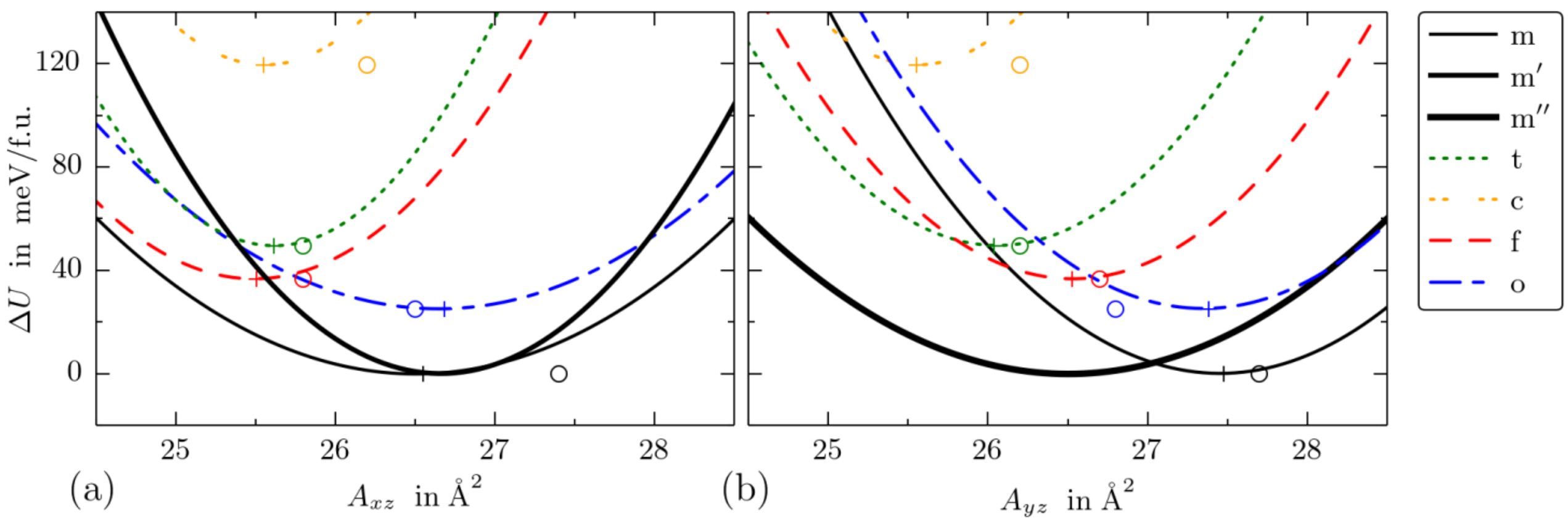

FIG. 5. (a) Calculated $\Delta U$ for all phases with plane strain in the *xz*-direction. Indicated are the *xz*-areas from data (o) compared to the *xz*-areas from the calculation (+). (b) Calculated $\Delta U$ for all phases with plane strain in the *yz*-direction. The curves m' represent the m-phase in the orientation chosen by Luo et al. [34] and others[47] while m refers to our choice in Table I. For comparison (b) contains the third possible orientation m" which stabilizes the m-phase for all possible strain conditions.

To obtain an impartial argument we have calculated the Helmholtz free energy $\Delta F$ for all strain planes *xy*, *xz,* and *yz,* and all phases. The results can be found in the SI[32], FIG. S3.

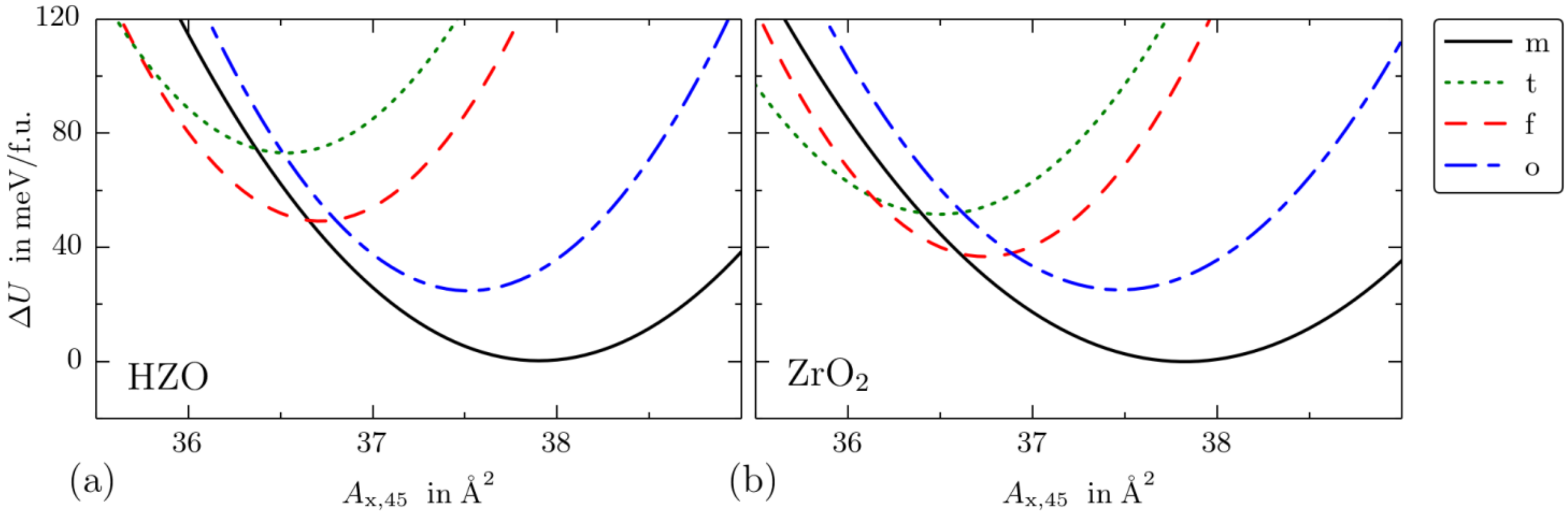

FIG. 6. (a) Calculated $\Delta U$ for m-, f-, t-, and o-phases of HZO with plane strain in the 45° rotated plane. (b) Calculated $\Delta U$ $ZrO_2$ with plane strain in the 45° rotated plane.

The conclusion is that a strain induced f-phase is only stable for a compressive strain in the xz-plain when the m-phase orientation of TABLE I is suppressed for some reason (see FIG. 5a). Unfortunately, for such an orientation the polarization direction is within the thin film plane. Therefore the grains can not be polarized by an external electric field perpendicular to the thin film plane and an existing polarization can not be detected.

So far the strain was oriented in the x, y, and z directions only. Tipping the *c*-axis out of the plane so that a measurable polarization could remain was suggested by Park[16]. In FIG. 6 we have calculated how an angle of 45° affects the phase stability by rotating the stress tensor around the x-axis. No distinction between m-phase and m'-phase is necessary since both orientations fall upon each other. A window of stability opens for the f-phase (compared to FIG. 5) but the crossover points to the m-phase are located at a corresponding stress of 8 GPa for HZO and 7 GPa for $ZrO_2$. Based on the high corresponding stress values for crossover points indicated by our calculations we conclude that the film stress is unlikely the most important factor to cause a phase stabilization of the f-phase with a metastable polarization along the direction perpendicular to the thin film plane.

The remaining question is whether the calculations are consistent with the conditions of the first experimental observation of the f-phase in Mg-PSZ (Mg doped, partially stabilized zirconium) by Kisi [18]. The material was described as a matrix of Mg-stabilized c-phase containing grains in the t-phase. Upon cooling, the cubic matrix thermally contracts uniformly, whereas the *c*-axis of the embedded t-phase grains contracts faster, the *a*-axis and *b*-axis slightly slower. This results in a tensile stress condition on the t-phase grains in the *c*-direction of about 3.6 GPa and a small compressive stress in the a- and b-direction of about 0.5G Pa.

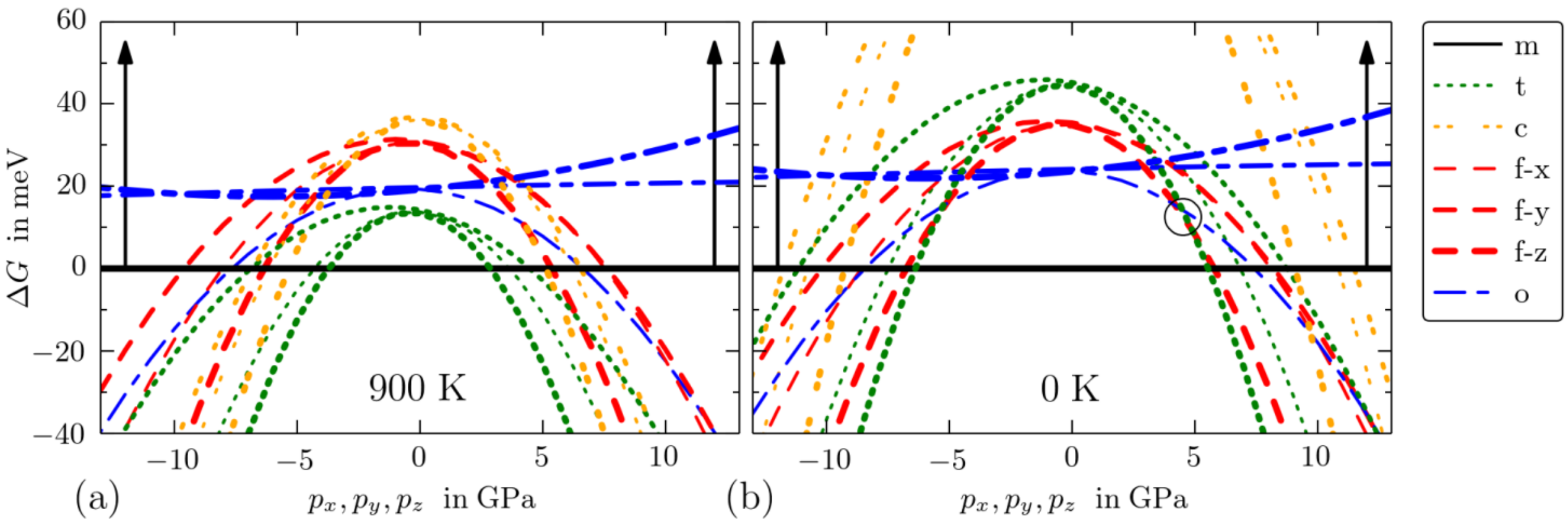

FIG. 7. (a) Calculated $\Delta G$ relative to the m-phase for all phases and all unidirectional stress directions x, y and z for $ZrO_2$ at 900 K. The included total energies are from the pure oxide without effect of dopant. (b) Calculated $\Delta G$ for $ZrO_2$ at 0 K. The arrows indicate that the m-phase is destabilized in experiments due to Mg-doping [18]. Stable range of f-phase indicated by (o).

By using the volume $V_0$ of the zero stress cell, the stress tensor $\sigma_{ij}$, and strain tensor $s_{ij}$ we have calculated the anisotropic Gibbs energy as

$$G_i = F + V_0\sigma_{ii}s_{ii} = U - TS + V_0\sigma_{ii}s_{ii}\,, \quad i = x,\ y,\ z \qquad (4)$$

and the Gibbs energy differences $\Delta G_i$ relative to the m-phase for all phases and stress directions, keeping the stress in the orthogonal directions as zero. The results in FIG. 7 show the pressure dependent Gibbs energy at 900 K and 0 K. Since in the experiment the Mg doping stabilized the enclosed grains in the t-phase at room temperature and above, we assume that the doping shifts the Gibbs energy $G$ of the m-phase above the Gibbs energy $G$ of the t-phase. Reducing the temperature shifts the t-phase up relative to the f-phase due to the entropy contribution $TS$. In FIG. 7 the f-phase is stabilized at a compressive stress. Our calculations neither included the effects of Mg dopants nor the compressive stress in $a$ and $b$ direction, which might open a larger window of stability for the f-phase. Therefore, we conclude that the computed results are in reasonable agreement with the experimental findings of a f-phase under unidirectional, tensile stress conditions. The major conclusion out of this section is that the f-phase with a measurable polarization in the normal direction cannot be stabilized with compressive film stress alone.

### E. Surface Energy and Energy Crossover

It is known that surface energy effects are responsible for size driven phase transformations on the nanoscale[17]. An example is the transformation of nanocrystalline $ZrO_2$ from the m- to the t-phase which is utilized to increase the capacitance of nowadays DRAM capacitors. The surface energies $\gamma$ of m- and t-$ZrO_2$[20,21] as well as m-$HfO_2$[23] have been measured, leading to a consensus at least about the range of the values between 2-4 J/m$^2$. Surface energies $\gamma$ have also been calculated for various orientations by Christensen[22] and range between 1.246-2.464 J/m$^2$ for m-$ZrO_2$ and 1.239-1.694 J/m$^2$ for t-$ZrO_2$, but the medium surface orientation of the grains is not known. TABLE V contains a collection of literature results. The surface energy introduces a further dependence in the Helmholtz free energy $F_\gamma$ (or the Gibbs energy $G_\gamma$, when the pressure is included) and the phase stability of nanocrystals of surface area $\Omega$ is determined from

$$F_\gamma = F + \gamma\,\Omega = U - TS + \gamma\,\Omega \qquad (5)$$

The requisite for the existence of t-$ZrO_2$ nanocrystallites is a smaller surface energy compared to the m-phase, which is confirmed by the data in TABLE V and consistent with a size driven phase transformation and energy crossover below a diameter of about 24 nm[62]. In the case of $HfO_2$ the energy crossover to the t-phase has been observed below about 3 nm [63]. The surface energy is also dependent on the termination with H or OH (anhydrous or hydrous) or other radicals. As a consequence, an exposure of tetragonally stabilized $ZrO_2$ with water could lead to a transformation back to the m-phase, an effect known as low-temperature degradation[64]. Finally, the surface energies have been measured in powder whereas many applications use nanocrystalline thin films with additional effects from interface energy. The essential observation in the size driven m→t transformation in $ZrO_2$ compared to $HfO_2$ is that in $HfO_2$ a much larger surface area to volume ratio is required. This implies a smaller difference between the surface energy of the m- and t-phase in $HfO_2$ compared to $ZrO_2$.

TABLE V. Surface energy $\gamma$ of $Hf_{1-\chi}Zr_{\chi}O_2$ for the competing phases from data calculation if available and from our model (bold).

| | $\gamma$ ($HfO_2$) [J/m$^2$] | $\gamma$ (HZO) [J/m$^2$] | $\gamma$ ($ZrO_2$) [J/m$^2$] |
|---|---|---|---|
| m-phase, model | 3.4 | 3.2 | 3.0 |
| data (anhydrous) | 3.7±0.1[23] | | 3.45±0.28[65] |
| data (hydrous) | 2.8±0.1[23] | | 2.86±0.31[65] |
| data | | | 6.4±0.2[20] |
| calculation | - | | 1.246-2.464[22] |
| o-phase, model | 3.3 | 2.9 | 2.5 |
| f-phase, model | 3.15 | 2.575 | 2.0 |
| t-phase, model | 3.1 | 2.5 | 1.9 |
| data (anhydrous) | | | 1.03±0.05[65] |
| data (hydrous) | | | 1.23±0.04[65] |
| data | | | 2.1±0.05[20] |
| calc | - | - | 1.239-1.694[22] |
| c-phase, model | 3.05 | 2.425 | 1.8 |

Since our question concerns the stability of the f-phase we need values for the surface energy of all the competing phases which have neither been measured nor calculated. Nonetheless, we need values for the composition dependence since the ferroelectric phase is observed in HZO. We chose the missing values as model parameters. The choice has been guided by the existing data, most importantly by the smaller difference of $\gamma$ (m-$HfO_2$) and $\gamma$ (t-$HfO_2$) compared to $\gamma$ (m-$ZrO_2$) and $\gamma$ (t-$ZrO_2$). Additionally, we decrease $\gamma$ with increasing crystal symmetry. The composition dependence finally is modeled from linear interpolation

$$\gamma\,(Hf_{1-\chi}Zr_{\chi}O_2) = (1-\chi)\,\gamma\,(HfO_2) + \chi\,\gamma\,(ZrO_2) \quad (6)$$

After our choice of values for the surface energy as given in TABLE V we now have a complete model to calculate the phase stability of $Hf_{1-\chi}Zr_{\chi}O_2$ and are able to compare with existing thin film data. Assuming the absence of a pressure or strain constraint we determine the stable phase from the minimum of the Helmholtz free energy including the surface contribution $F_\gamma = U - TS + \gamma\ \Omega$. The total energy $U$ and the entropy contribution $TS$ are calculated fundamentally from DFT as above. At this point it is clear that these final results depend on the accuracy of the total energy values and entropy contributions, justifying the lengthy discussion above. The surface contribution $\gamma\Omega$ finally uses a phase and composition dependent model parameter $\gamma$ and a geometry model for the calculation of the surface area $\Omega$, assuming a cylinder surface with height $h$ from film thickness and radius $r=h/2$, if not given otherwise.

We continue with a comparison of the model results with data. TABLE VI contains the calculated results for a 9 nm thin film at 80 K, 300 K and a 900 K anneal temperature modeling the experimental results of Müller [11]. For all temperatures the 9 nm thick film is calculated to be monoclinic even at anneal temperatures. For smaller crystallites Shandalov[63] observed a phase transformation between a size of 7.6 nm and 3.1 nm. For 3.1 nm crystallites we calculate that f-$HfO_2$ has the lowest Helmholtz free energy in close competition with t-$HfO_2$. Since the crystallites have been annealed we have calculated the values at 900 K and find t-$HfO_2$ with the lowest Helmholtz free energy. We hypothesize that such crystallites can be transformed to f-$HfO_2$ under the influence of an E-field. Finally, we calculate the values for crystallites of 2 nm diameter observed by Bohra et al. [66]. At this small size we clearly find f-$HfO_2$ to have the lowest Helmholtz free energy. Bohra et al. identified the crystallites to be of o-phase from an electron diffraction pattern. A possible explanation is, that for extremely small crystallites a further model is required which adds the effect of size induced hydrostatic pressure [67 68].

The next results in TABLE VI model the 9 nm thin film HZO from Müller et al. at 80 K, 300 K and 900 K. At 80 K f-HZO is stable with an energy of 7 meV/f.u. below the competing t-HZO. At 300 K the f-HZO is still stable but only 1 meV/f.u. lower than the competing t-HZO. At anneal temperature the t- and c-phase are more favorable, and a phase transformation to the f-phase has to take place upon cooling. The calculations for a 7 nm film show the tendency of a thinner film to become tetragonal. The data for a 7 nm film from Müller [2] compared to a 9 nm film show this tendency in a rudimentary antiferroelectric hysteresis curve. This size driven transformation from t- over f- to m-HZO is shown in the calculation and data of Park [16] who investigated a thickness series of 5.5 nm, 9.2 nm, 14 nm, 19 nm, and 24 nm.

The last group of results in TABLE VI is about $ZrO_2$ starting again with a model for the data from Müller [11]. The 9 nm film is paraelectric at 80 K and room temperature. The antiferroelectric character becomes visible in an electric field driven paraelectric to ferroelectric phase transformation. The strength of the required electric field in Müller's data decreases from about 1 MV/cm to about 0.5 MV/cm when lowering the temperature from 230 K to 80 K. In the calculation the film shows to be tetragonal in both cases, but the energy difference to the f-phase lowers from 10 meV/f.u. to 5 meV/f.u. which is consistent with a decrease of the required electric field. The last data in the table concern the crossover size from the t- to the m-phase. According to the data of Yashima [62] a diameter of 24 nm is sufficient whereas in the model a size of 36 nm is necessary to yield the m-phase to have the lowest energy.

TABLE VI. Comparison of $\Delta F_{\gamma}$ relative to the m-phase for $HfO_2$ crystallites of different size and temperature in comparison with data. Bold numbers point out the lowest Helmholtz free energy according to our model.

| $T$ [K] | $h/r$ [nm] | calculated $\Delta F_{\gamma}$ in meV/f.u. m | o | f | t | c | observed phase |
|---|---|---|---|---|---|---|---|
| | | | | $HfO_2$ | | | |
| 300 | bulk | 0 | 23 | 59 | 78 | 113 | m-phase[3] |
| 80 | 9/4.5 | 0 | 11 | 30 | 50 | 83 | m-phase[11] |
| 300 | 9/4.5 | 0 | 11 | 29 | 42 | 70 | m-phase[11] |
| 900 | 9/4.5 | 0 | 8 | 24 | 16 | 28 | m-phase[11] |
| 300 | 7.6/8.0 | 0 | 13 | 34 | 48 | 78 | m-phase[63] |
| 300 | 3.1/2.4 | 0 | -6 | -13 | -9 | 12 | t-phase[63] |
| 900 | 3.1/2.4 | 0 | -9 | -18 | -35 | 30 | t-phase[63] |
| 300 | 2.0/1.0 | 0 | -36 | -88 | -98 | -92 | o-phase[66] |
| | | | | HZO | | | |
| 300 | bulk | 0 | 24 | 48 | 58 | 95 | - |
| 80 | 9/4.5 | 0 | -12 | -28 | -21 | 19 | f-phase[11] |
| 300 | 9/4.5 | 0 | -12 | -28 | -27 | 1 | f-phase[11] |
| 900 | 9/4.5 | 0 | -13 | -30 | -51 | -43 | t-phase[11] |
| 300 | 7.0/3.5 | 0 | -26 | -57 | -59 | -35 | f,t-phase[2] |
| 300 | 5.5/3.0 | 0 | -36 | -78 | -83 | -61 | t-phase[16] |
| 300 | 9.2/4.6 | 0 | -14 | -32 | -31 | -4 | f-phase[16] |
| 300 | 14/7 | 0 | -1 | -4 | -1 | 30 | f-phase[16] |
| 300 | 19/9.5 | 0 | 5 | 9 | 15 | 47 | m-phase[16] |
| 300 | 24/12 | 0 | 9 | 17 | 24 | 57 | m-phase[16] |
| | | | | $ZrO_2$ | | | |
| 300 | bulk | 0 | 24 | 34 | 37 | 72 | m-phase[3] |
| 80 | 9/4.5 | 0 | -41 | -95 | -100 | -63 | t-phase[11] |
| 300 | 9/4.5 | 0 | -41 | -96 | -106 | -84 | t-phase[11] |
| 900 | 9/4.5 | 0 | -43 | -99 | -129 | -137 | t-phase[11] |
| 300 | 24/12 | 0 | 0 | -15 | -17 | 13 | m-phase[62] |
| 300 | 36/18 | 0 | 8 | 1 | 1 | 33 | m-phase |

The results of the free energy model for the 9 nm film data series of Müller [11] in comparison with bulk free energy is shown in FIG. 8. $HfO_2$ is monoclinic for all temperatures. HZO is ferroelectric for room temperature and below. $ZrO_2$ is tetragonal for all temperatures. The ferroelectricity in HZO arises in the model from the linear dependence of the surface energy on the composition and the increase of the surface energy with crystal symmetry.

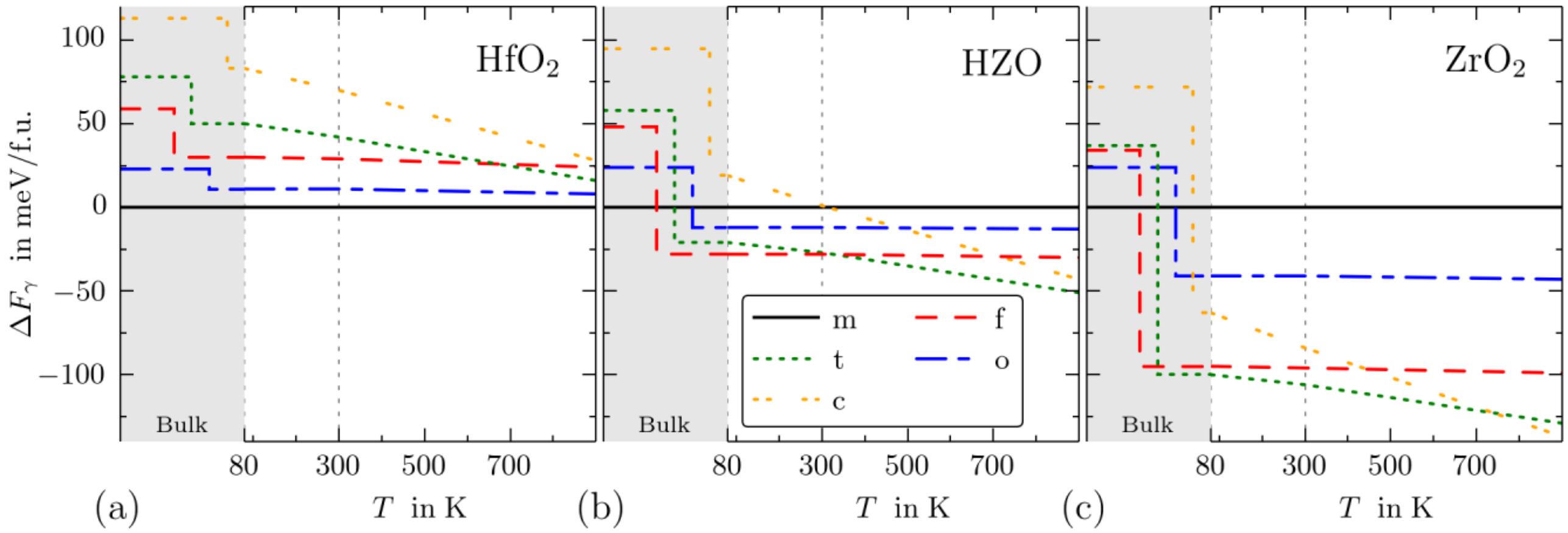

FIG. 8. Calculated Helmholtz free energy $\Delta F_\gamma$ for bulk (compare FIG. 3) and a 9 nm polycrystalline film for (a) $HfO_2$, (b) HZO, and (c) $ZrO_2$ above 80 K. The colored area guides the eye to the difference in Helmholtz free energy $\Delta F$ of bulk (intersection with the ordinate) and the 9 nm thin film at 80 K.

The conclusion from the comparison of the Helmholtz free energy model with available data is that the model describes the observations very well, although the involved energy differences are in the order of a few meV/f.u. This is possible with a model, where the error from the total energy contribution $U$ and the error from the entropy contribution $TS$ have been reduced to a small amount.

As a consequence, the model achieves a predictive capability. FIG. 9 shows the modeled Helmholtz free energy for the phases of $HfO_2$, HZO, and $ZrO_2$ relative to the phase with the lowest Helmholtz free energy at these conditions as a function of film thickness with cylindrical grains.

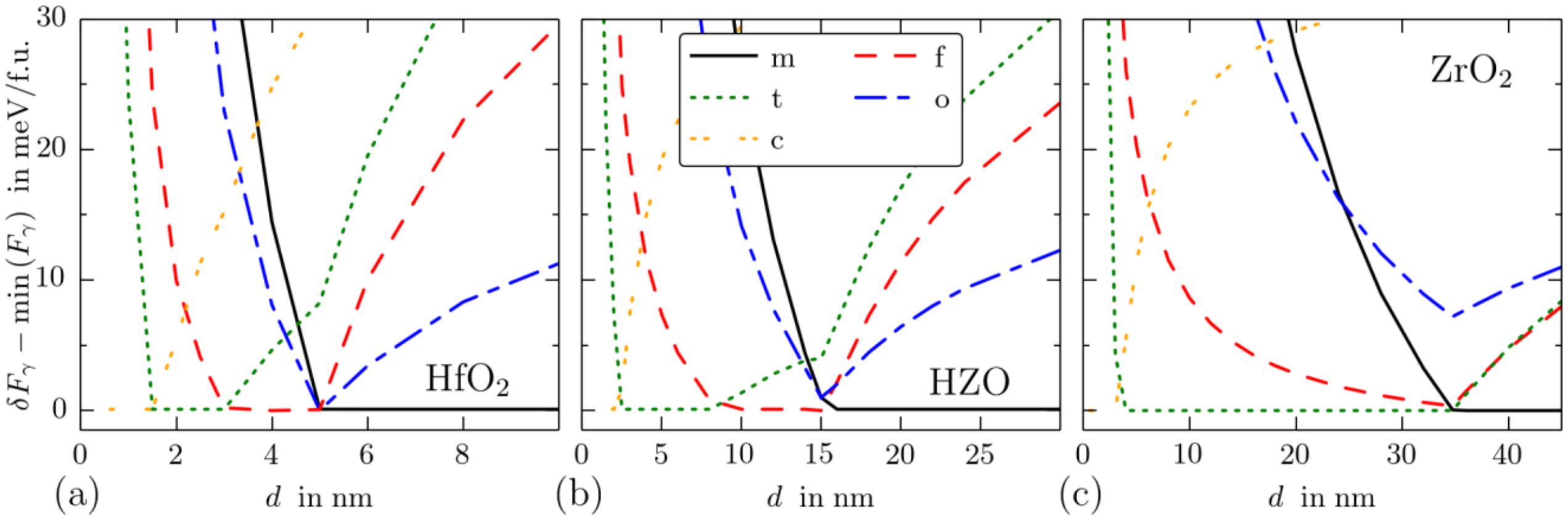

FIG. 9. Calculated free energy difference $\Delta F_\gamma$ relative to the phase with minimal Helmholtz free energy at 300 K as a function of film thickness for (a) $HfO_2$, (b) HZO and (c) $ZrO_2$. A window of stability for the f-phase arises for $HfO_2$ from 3 nm to 5 nm and in HZO from 8 nm and 16 nm thin film strength

The crossover of $HfO_2$ from the high symmetry phases to the low symmetric m-phase happens for film thickness below 5 nm. The model predicts ferroelectric $HfO_2$ for a grain size between 3 nm and 5 nm, for smaller grains the t-phase and c-phase. For HZO a similar crossover exists but is shifted to larger grain size. The ferroelectric films exist in the thickness regime between 8 nm and 16 nm. Thinner films or smaller grains occur in the t-phase. In addition, our model suggests the existence of a ferroelectric phase in epitaxial HZO below a critical film strength of about 5 nm. For $ZrO_2$, the large surface energy difference of the t-phase is dominant such that no f-phase is stable, although it is energetically close. The energy disadvantage to the t-phase is small so that it can be overcome with an electric field (see Sec. F). Despite being the second most stable phase in the bulk, the o-phase is never the most stable phase for any grain volume.

### F. Electric Enthalpy and Field Driven Phase Transformation

The antiferroelectric behavior observed in doped $HfO_2$ [1] and $ZrO_2$ [11] has been interpreted as a field induced phase transformation [19]. We examine the consistency of this statement with the Helmholtz free energy model containing a contribution for a polarized crystal in an electric field as described by Souza [69]. The Helmholtz free energy in a volume $V_0$ with the field energy contribution is

$$F_{\gamma,E} = F_\gamma - V_0 DE = U - TS + \gamma \Omega - V_0 DE \qquad (7)$$

when $D=D(E) = \varepsilon_r\varepsilon_0 E+P$ is the electric displacement as a function of the macroscopic electric field $E$ which has the polarization $P$ as a permanent contribution and $\varepsilon_r\varepsilon_0 E$ as an induced contribution where $\varepsilon_r$ is the dielectric tensor. The total energy is obtained from a structural relaxation of the supercell including the field term. An induced polarization lowers the energy while the field strain of the crystal increases $U$. A further effect is a modification of the entropy from the field strain shifted phonon modes. However, we have neglected this contribution.

We have calculated the Helmholtz free energy $F_E$ for $ZrO_2$ bulk in the m-, t- and f-phase and show the difference relative to the m-phase at zero E-field in FIG. 10 a-b. A field in *y*-direction orthogonal to the polarization direction does not lead to a phase transformation for realizable fields. A field in the *z*-direction parallel to the polarization lowers the Helmholtz free energy of the f-phase significantly by about 10 meV/f.u. for 1 MV/cm. For bulk $ZrO_2$ a field of about 4 MV/cm would be required to give the f-phase a Helmholtz free energy below the m-phase.

But for thin film $ZrO_2$ the Helmholtz free energy of the t- and f-phase is already below the m-phase. Here a much lower field of about 1 MV/cm is sufficient for the f-phase to obtain the lowest energy. We can reproduce the temperature dependence of the field driven phase transformation in 9nm thin films measured by Müller [11] by calculating the change of the entropy contribution with temperature. In FIG. 10c we show the Helmholtz free energy of the t- and f-phase for 80 K and 230 K. At 80 K a field strength of only 0.5 MV/cm is sufficient to induce the phase transformation.

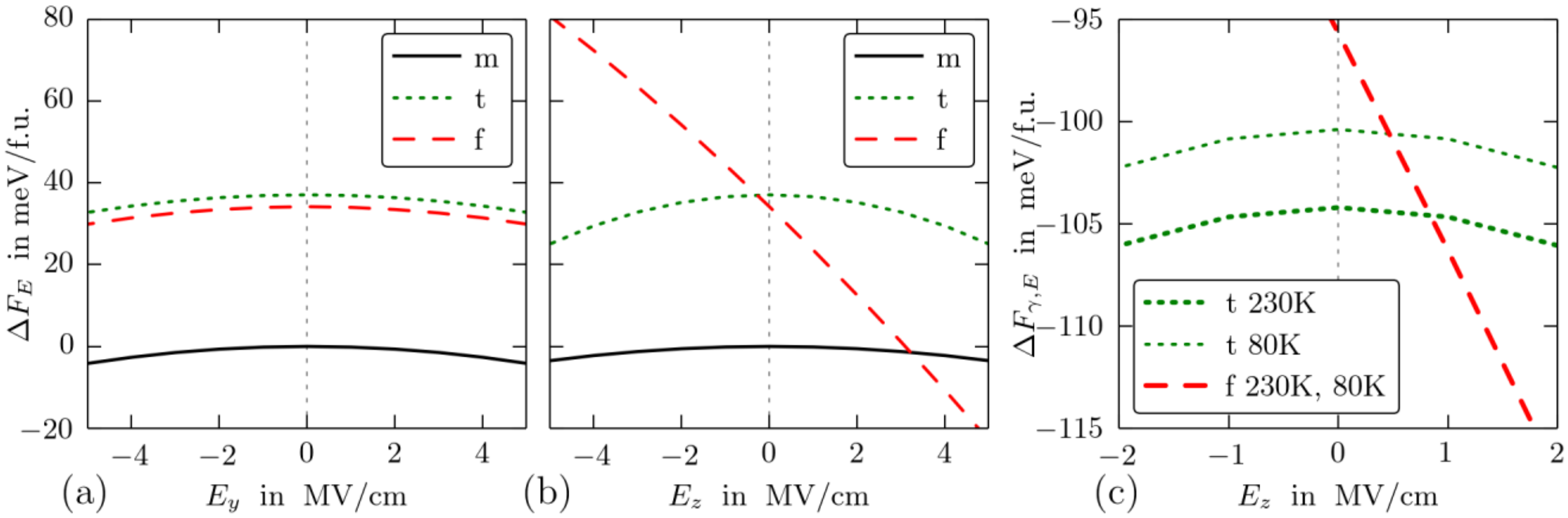

FIG. 10. Calculated Helmholtz free energy difference $\Delta F_{\gamma,E}$ for bulk $ZrO_2$ relative to the m-phase at zero E-field and $T$=300K, (a) for an E-field in $y$-direction, (b) for an E-field in $z$-direction and a polarization $P$ in negative $z$-direction. (c) Shift of the energy crossover in 9 nm $ZrO_2$ for the same E-field/Polarization configuration for a temperature increase from $T$=80K to $T$=230K. For the t-phase the Helmholtz free energy difference to the m-phase $\Delta F_{\gamma,E}$ is dependent on temperature, while for the f-phase it is nearly independent on temperature (see FIG. 3).

The necessary E-field for the field driven phase transformation from the t- to the f-phase was so far determined from the Helmholtz free energy difference $F_{\gamma,E}$(f-phase)- $F_{\gamma,E}$(t-phase) in an equilibrium state. A further requirement for a phase transformation is a sufficiently low energy barrier to the new Helmholtz free energy minimum. As a good approximation to the transition barrier[34] we have calculated the maximum of the transition path from the t- to the f-phase. We obtain a barrier of only 30 meV/f.u. for the chosen orientation in TABLE I. Such a small barrier should be overcome by thermal energy alone, hence removal of the external field relaxes the structure back to the t-phase.

Similar values have been recently found in the literature (30 meV/f.u. in Huan [49] and 30 meV/f.u. from Reyes-Lillo [47]). The low barrier is related to a soft phonon mode interpolating between t- and f-phase. Reyes-Lillo et al. found a total energy difference of only 1 meV between f- and t-phase and calculated a critical field strength of 1.2 MV/cm necessary to overcome the barrier of 30 meV/f.u. They do not explain why the f-phase relaxes back into the t-phase, after the field strength is reduced to zero.

To show the consistency of the Helmholtz free energy calculation we have extracted the dielectric constants $\varepsilon_{ri}$, $i=x,y,z$ (diagonal elements of the dielectric tensor) and the polarization $P$ from a second order fit to the computed values of the field energy $V_0\, DE= V_0(\varepsilon_r\varepsilon_0 E+P)E$. TABLE VII shows the comparison of these values $\varepsilon_{E\text{-}field}$ for m-, f- and t-$HfO_2$ with values $\varepsilon_{ionic}$ from a conventional linear response calculation to an ionic perturbation and $\varepsilon_{total}$ including the electronic contribution. The results only show a small deviation between $\varepsilon_{E\text{-}field}$ and $\varepsilon_{ionic}$ indicating a slightly nonlinear behavior.

TABLE VII. Diagonal elements of the dielectric tensor and averaged dielectric constants calculated from the Helmholtz free energy and from a linear response calculation.

| | HfO2 | | | HZO | | ZrO2 | |
|---|---|---|---|---|---|---|---|
| | $\varepsilon_{E\text{-}field}$ | $\varepsilon_{ionic}$ | $\varepsilon_{total}$ | $\varepsilon_{ionic}$ | $\varepsilon_{total}$ | $\varepsilon_{ionic}$ | $\varepsilon_{total}$ |
| m-phase P14 | | | | | | | |
| x | 20.0 | 19.8 | 24.7 | 21.5 | 26.7 | 23.3 | 29.0 |
| y | 18.0 | 18.1 | 22.9 | 19.6 | 24.8 | 21.3 | 27.0 |
| z | 15.0 | 15.0 | 19.5 | 15.6 | 20.5 | 16.3 | 21.5 |
| Σ/3 | 17.7 | 17.7 | 22.4 | 18.9 | 24.1 | 20.3 | 25.8 |
| o-phase P61 | | | | | | | |
| x | - | 20.3 | 25.3 | - | | 23.7 | 29.5 |
| y | - | 18.0 | 22.8 | - | | 21.6 | 27.3 |
| z | - | 15.5 | 20.1 | - | | 16.8 | 22.1 |
| Σ/3 | - | 17.9 | 22.7 | - | | 20.7 | 26.3 |
| f-phase P29 | | | | | | | |
| x | 24.0 | 23.8 | 28.8 | 26.0 | 31.5 | 28.4 | 34.4 |
| y | 19.0 | 19.3 | 24.2 | 20.7 | 25.9 | 22.1 | 27.8 |
| z | 26.0 | 23.1 | 28.0 | 24.6 | 29.9 | 26.1 | 31.7 |
| Σ/3 | 23.0 | 22.1 | 27.0 | 23.8 | 29.1 | 25.5 | 31.3 |
| t-phase P137 | | | | | | | |
| x,z | 55.0 | 51.8 | 56.9 | 54.3 | 60.4 | 55.4 | 61.5 |
| y | 20.0 | 19.2 | 24.0 | 19.9 | 25.3 | 19.9 | 25.3 |
| Σ/3 | 43.3 | 40.9 | 45.9 | 42.8 | 48.7 | 43.6 | 49.4 |
| c-phase P225 | | | | | | | |
| x,y,z | - | 30.9 | 36.0 | - | | 41.5 | 47.6 |
| Σ/3 | - | 30.9 | 36.0 | - | | 41.5 | 47.6 |

The comparison of the extracted polarization $P_{E\text{-}field}$ for f-$HfO_2$ with values $P_{Berry}$ from a conventional Berry phase calculation is shown in TABLE VIII. The values match excellent. Furthermore, the table shows the dependence of the polarization $P$ of $Hf_{1-\chi}Zr_{\chi}O_2$ on the stoichiometry $\chi$. The polarization increases with Zr content. This can be understood by the increasing values of the Born charges obtained from the linear response calculation. TABLE VIII shows only the diagonal values of the Born charge tensor. A simplified estimation based on equation (1) including the diagonal values and a perturbation in z-direction around the centrosymmetric configuration leads to a good estimation and explains the stoichiometric trend of the polarization.

TABLE VIII. Polarization

| | HfO2 | | | HZO | | ZrO2 | |
|---|---|---|---|---|---|---|---|
| | $P_{E\text{-}field}$ | $P_{Berry}$ | | $P_{Berry}$ | | $P_{Berry}$ | |
| z | 0.505 | 0.504 | | 0.541 | | 0.579 | |
| | $\varepsilon_{E\text{-}field}$ | $\Delta z$ | $Z$ | $\Delta z$ | $Z$ | $\Delta z$ | $Z$ |
| Zr | 20.0 | 0.003 | 5.08 | 0.002 | 5.23 | 0.002 | 5.85 |
| O1 | 18.0 | -0.138 | -2.53 | -0.135 | -2.54 | -0.133 | -2.58 |
| O2 | 15.0 | 0.008 | 2.52 | 0.008 | 2.58 | 0.006 | 2.60 |
| eq. (1) | 17.7 | 0.67 | | 0.69 | | 0.72 | |

## IV. CONCLUSION

In this paper we have investigated the origin of the ferroelectricity in $Hf_{1-\chi}Zr_{\chi}O_2$ for HZO with $\chi = 0.5$, and the origin of antiferroelectricity for $ZrO_2$ with DFT calculations, and a phenomenological model for the surface energy contribution. From total energy calculations we found that the bulk ferroelectric phase $Pca2_1$ is not stable. The computations for several polymorphs were compared to a number of literature results to estimate the theoretical uncertainty and find the most reliable method. We decided for the LDA functional and norm-conserving TM pseudopotentials. The temperature dependent Helmholtz free energy was calculated accordingly. As a first possible mechanism to stabilize the ferroelectric phase we investigated the film strain. Although we could reproduce all relevant stress and strain dependent phenomena in $HfO_2$ and $ZrO_2$ with our model, we only found a small window of rather high compressive film strain to allow a stabilization of the ferroelectric phase. Since the thus stabilized structure has a polarization in the strain plane, which is unobservable, we found the stress/strain mechanism unlikely to be the major cause of the stabilization. As a second possible mechanism we investigated the surface energy contribution to the Helmholtz free energy. Since the surface energy of ferroelectric grains has not been measured or calculated so far we build a phenomenological model for all polymorphs and all stoichiometry values by interpolating between existing values using decreasing values for increasing crystal symmetry. With the Gibbs- / Helmholtz free energy model containing an ab initio computed part for total energy, entropy, stress or strain, and a phenomenological part for the surface energy we could reproduce the observed phases for nanosized thin films and grains and found stable ferroelectric HZO in a size window around 10 nm at room temperature in absence of strain. Based on the model we furthermore predict a similar ferroelectric window in $HfO_2$ around 4 nm and below 5 nm in epitaxial HZO. For $ZrO_2$ no stable ferroelectric nanosized grains exist. After calculating the field dependent contribution to the Helmholtz free energy self-consistently we found an electric field of about 1 MV/cm sufficient to give ferroelectric grains of 9 nm at room temperature the lowest Helmholtz free energy. Furthermore, reducing the temperature favors the stability and decreases the crossover field in accordance with experimental data.

## ACKNOWLEDGMENTS

The author wants to thank U. Schroeder and T. Schenk from NamLab and U. Böttger and S. Starschich from RWTH Aachen for discussions. The German Research Foundation (Deutsche Forschungsgemeinschaft) is acknowledged for funding this research in the frame of the project “Inferox” (project no. MI 1247/11-1). The authors gratefully acknowledge the Gauss Centre for Supercomputing e.V. (www.gauss-centre.eu) for funding this project by providing computing time on the GCS Supercomputer SuperMUC at Leibniz Supercomputing Centre (LRZ, www.lrz.de).